\documentclass[ %
  aip,
 amsmath,amssymb,
 reprint,%
]{revtex4-1}

\usepackage{graphicx}
\usepackage{dcolumn}
\usepackage{bm}

\usepackage[utf8]{inputenc}
\usepackage[T1]{fontenc}
\usepackage{mathptmx}
\usepackage{etoolbox}

\usepackage{changes}
\usepackage{xcolor}
\definechangesauthor[name={}, color=red]{anon}

\makeatletter
\def\@email#1#2{%
 \endgroup
 \patchcmd{\titleblock@produce}
  {\frontmatter@RRAPformat}
  {\frontmatter@RRAPformat{\produce@RRAP{*#1\href{mailto:#2}{#2}}}\frontmatter@RRAPformat}
  {}{}
}%
\makeatother
\begin{document}

\preprint{AIP/123-QED}

\title[EM Condensation and Critical Phenomena in LJ fluid]{Eigen Microstate Condensation and Critical Phenomena in the Lennard-Jones Fluid}

\author{Lan Yang}
\altaffiliation[These authors ]{contributed equally to this work.}
\affiliation{School of Systems Science $\&$ Institute of Nonequilibrium Systems, Beijing Normal University, Beijing 100875, China}

\author{Zhaorong Pang}
\altaffiliation[These authors ]{contributed equally to this work.}
\affiliation{Arnold-Sommerfeld Center for Theoretical Physics, Theresienstr, 37, Munich 80333, Germany}

\author{Chongzhi Qiao}
\altaffiliation[Electronic mail: ]{chongzhi.qiao@bnu.edu.cn}
\affiliation{School of Systems Science $\&$ Institute of Nonequilibrium Systems, Beijing Normal University, Beijing 100875, China}

\author{Gaoke Hu}
\affiliation{College of Physics, Nanjing University of Aeronautics and Astronautics, Nanjing, 211106, China}
\affiliation{China Key Laboratory of Aerospace Information Materials and Physics (NUAA), MIIT, Nanjing 211106, China}

\author{Jiaqi Dong}
\affiliation{Lanzhou Center for Theoretical Physics and Key Laboratory of Theoretical Physics of Gansu Province, Lanzhou University, Lanzhou, Gansu 730000, China}

\author{Rui Shi}
\altaffiliation[Electronic mail: ]{ruishi@zju.edu.cn}
\affiliation{Zhejiang Key Laboratory of Micro-nano Quantum Chips and Quantum Control, School of Physics, Zhejiang University, Hangzhou 310027, China}

\author{Xiaosong Chen}
\altaffiliation[Electronic mail: ]{chenxs@zju.edu.cn}
\affiliation{School of Systems Science $\&$ Institute of Nonequilibrium Systems, Beijing Normal University, Beijing 100875, China}
\affiliation{Institute for Advanced Physics, Zhejiang University, Hangzhou 310027, China}

\date{\today}

\begin{abstract}
Despite extensive study of the liquid-gas phase transition, accurately determining the critical point and the critical exponents in fluid systems through direct simulation remains a challenge.
We employ the eigen microstate theory (EMT) to investigate the liquid-gas continuous phase transition in Lennard-Jones (LJ) fluid within the canonical ensemble.
In EMT, the probability amplitudes of eigen microstates serve as the order parameter. Using finite-size scaling of probability amplitudes, we simultaneously determine the critical temperature, $T_c = 1.188(2)$, and critical density, $\rho_c = 0.320(4)$.
Furturemore, we obtain critical exponents of the LJ fluid, $\beta = 0.32(2)$ and $\nu = 0.64(3)$, which demonstrate a great agreement with the Ising universality class.
This method also reveals the mesoscopic structure of the emergent phase, characterizing the three-dimensional (3D) spatial configuration of the fluid in the critical region.
This work also confirms the finite-size scaling behavior of the probability amplitudes of the eigen microstates in the critical region.
The EMT provides a powerful tool for studying the critical phenomena of complex fluid system. 

\end{abstract}

\maketitle

\section{\label{sec:1}Introduction}
The liquid-gas phase transition plays a crucial role in numerous scientific fields and industrial processes.	
For instance, liquid-gas phase transitions form the basis of various industrial operations, including distillation, flash evaporation, and adsorption.	
Distillation alone accounts for approximately half of the energy used in separation processes and over 20\% of the total energy consumption in the United States~\cite{sholl2016seven}.	
Therefore, understanding the mechanism of liquid-gas phase transitions is critical for process optimization and reducing energy consumption.

The Lennard-Jones (LJ) potential, which incorporates a short-range repulsive force and a long-range attractive force, is a widely used model for simulating liquid-gas phase transitions and is considered crucial for studying the behavior of simple fluids~\cite{1924,LJ_1931,Wood1957,Rahman1964}. 
It has been employed to explore a range of phenomena, including liquid-gas and liquid-liquid phase transitions~\cite{rovereGLTransition2D1990,Wilding1998,watanabe2012a,Sheibani2020}, melting and vitrification~\cite{Berlin2013}, fluid behaviors in confined nanoscale environments~\cite{Zhang2011,qiao2020enhancing}, porous nanomaterials~\cite{Alain2000a,liu2015high,qiao2019connect}, and transport properties~\cite{zhao2017surface,Geada2018,LAUTENSCHLAEGER2019}. 
In studies involving complex fluid perturbation theories, the LJ potential is frequently adopted as a reference fluid~\cite{johnsonLJEquationState1993,wang2023microstructure,wang2024molecular}.

Since the advent of computer simulations, a wide range of techniques has been applied to investigate the liquid-gas phase transition, including Monte Carlo simulations, molecular dynamics simulations, and classical density functional theory.
Panagiotopoulos~\cite{panagiotopoulos1987direct,panagiotopoulos1988phase} introduced the Gibbs ensemble Monte Carlo method to study phase transitions in fluid systems. 
This approach employs two separate simulation boxes to represent the liquid and gas phases under constant volume, temperature, and chemical potential conditions.
However, near the critical point, significant fluctuations in particle number and system volume make the Gibbs ensemble Monte Carlo approach unstable~\cite{panagiotopoulos2000monte}.
Watanabe utilized the direct molecular dynamics simulation to construct liquid-gas interface and determined the densities of the gas and liquid phases by fitting the system's density profile near the interface~\cite{watanabe2012a}.
By applying the Binder cumulant and finite-size scaling analysis, the critical point and critical exponents were obtained. 
Wilding~\cite{wilding1995critical} used mixed-field finite-size scaling and histogram reweighting techniques to determine the critical parameters of the LJ fluid.
More recently, Sammüller et al.~\cite{sammuller2025neural} combined neural density functional theory with finite-size scaling to identify the LJ fluid's critical point.
Despite substantial efforts to determine the critical properties of the LJ fluid, several fundamental questions remain unanswered.
Can the mesoscopic spatial structure of the liquid-gas critical point be resolved?
Is it possible to determine the LJ fluid's critical point without relying on higher-order moments, Binder cumulants, or knowledge derived from the Ising model?

Recently, the eigen microstate theory (EMT) has introduced a novel approach to addressing this challenge~\cite{Hu2019,Sun2021}. 
This method, which relies on microstates appropriately selected from experiments or simulations, represents the statistical ensemble over the observation period as a normalized matrix. 
From this matrix, eigen microstates are derived. Consequently, any microstate can be expressed as a linear superposition of these eigen microstates, with the corresponding eigenvalues acting as probability amplitudes. 
The dominant eigenvalue can then be interpreted as the order parameter of the phase transition, thereby enabling more direct analysis of complex systems.
The EMT has been effectively employed to investigate the equilibrium~\cite{Hu2019,Sun2021,Zhang2018} and the non-equilibrium~\cite{Li2021,zhang2023eigenstates} phase transition. Also other complex systems~\cite{Li2016}, such as the Earth systems~\cite{Fan2021,WANG2024,chen_2024,Chen2021}, the quantum Rabi model~\cite{Hu2023}.

In this study, we apply EMT to investigate the critical point and critical exponents of the LJ fluid. 
We employ canonical ensemble (NVT) Monte Carlo simulations to generate microstates of the LJ fluid. 
We present the evolution of the first eigen microstates across a range of temperatures. 
The mesoscopic structure of the emergent phase is characterized by the eigen microstates, which capture the spatial structure of the fluid in the critical region.
Finite-size scaling analysis is indispensable for studying critical phenomena in finite systems. By combining EMT with finite-size scaling~\cite{Kadanoff1966} of probability amplitudes, we determine critical density and critical temperature, simultaneously.
Crucially, this methodology eliminates the need for higher-order moments or Binder cumulants to identify critical points. 
Critical exponents $\beta$ and $\nu$ are derived through finite-size scaling of probability amplitudes. 
Our results demonstrate strong concordance with the three-dimensional Ising universality class and confirm the finite-size scaling law for probability amplitudes in the critical region.

This paper is organized as follows: the details of the Monte Carlo simulation, eigen microstate theory for the LJ fluid system, and the finite-size scaling of probability amplitudes are briefly introduced in Sec.~\ref{sec:2}. In Sec.~\ref{sec:3}, With the help of the finitie-size scaling of probability amplitudes, the critical temperature, $T_c$, critical density, $\rho_c$, and exponents, $\beta$ and $\nu$, are obtained. We further confirm the scaling relation in the critical region. 
We also present the spatial structure of the first two eigen microstates, $U_1$ and $U_2$, across various temperatures. These results characterize the mesoscopic structural evolution throughout the phase transition, a feature that is otherwise obscured in the raw microstates.
In the final section, we summarize our findings and outline potential directions for further studies.

\section{\label{sec:2}Eigen microstates and phase transitions}

\subsection{\label{sec:2.1} Eigen Microstates of the LJ fluid}
In this work, the Canonical ensemble Monte Carlo simulation is adopted to generate the microstates of the LJ system around critical point. 
A cubic box of size $L$ is used to simulate the LJ fluid.
The interaction between particles $i$ and $j$ is,
\begin{equation}
    \phi(r_{ij}) = 4\epsilon \left[ \left( \frac{\sigma}{r_{ij}} \right)^{12} - \left( \frac{\sigma}{r_{ij}} \right)^{6} \right], \label{eq:LJpotential}
\end{equation}
where $r_{ij}$ is the distance between particles $i$ and $j$, $\epsilon$ is the well depth, and $\sigma$ is the diameter.
The interaction potential is truncated at a distance $R_c=2.5\sigma$, \cite{wilding1995critical,vaporliquid1991}
\begin{equation}
    \phi_{\mathrm{tr}}(r) = \left\{
    \begin{array}{ll}
        \phi(r) & \text{for } r \leq R_c \\
        0 & \text{for } r > R_c
    \end{array}
    \right. .
    \label{eq:unshifted}
\end{equation}
In this work, the temperature, $T$, is measured in unit of $\epsilon/k_B$, where $\epsilon$ is the well depth of the LJ potential and $k_B$ is the Boltzmann constant.

To define the microstate in fluid system, the system is divided into $N_\sigma ={L_\sigma}^3=L^3/ \sigma^3$ sites, each site presents a small cube with length $\sigma$.
The local density of $i$th site is used to define the microstate, which is given by,

\begin{equation}
    \rho_i = \frac{n_i}{\sigma^3},
\end{equation}

where $n_i$ is the number of particles in the $i$th site.
The reduced density fluctuation in each site is given by, 

\begin{align}
\delta\rho_{i} = \frac{\rho_i - {\rho}}{{\rho}},
\end{align}
where the $\rho$ represents the density of the system, $\rho = N / L^3$, and $N = \sum_{i=1}^{N_\sigma} n_i$ is the number of particles in the whole system. The relation between number of particles $N$ and the number of sites $N_\sigma$ is given by $N_\sigma = N / (\rho\sigma^3)$.
The configuration of $j$th microstate can be presented as a vector with $N_\sigma$ components,
\begin{align}
    \delta\bm{\rho}(j) = 
        \left[
        \begin{array}{cccc}
            \delta\rho_1(j)\\
            \delta\rho_2(j)\\
            \vdots\\
            \delta\rho_{N_\sigma}(j)
        \end{array}
        \right].
\end{align}

With the collected microstates, a statistical ensemble can be constructed that characterizes the complex system. To construct the ensemble matrix, $M$ microstates are obtained from an equilibrated system at a given temperature. Typically, $M$ is chosen to be considerably larger than $N_\sigma$. This ensemble can be represented by an $N_\sigma \times M$ matrix $\bm{A}$, with its elements defined as,
\begin{equation}
    \bm{A}_{ij} = \frac{\delta\rho_i(j)}{\sqrt{C_0}},
\end{equation}
where $C_0 $ is the variance of the density fluctuations,

\begin{equation}
    C_0 = \sum_{i=1}^{N_\sigma} \sum_{j=1}^{M} \delta\rho_i(j)^2.
\end{equation}

The correlation between $i$th site and $j$th site can be considered as the inner product of the configuration of two sites, 

\begin{equation}
    \bm{K}_{ij} =\sum_{k=1}^M \delta\rho_{i}(k) \delta\rho_{j}(k).
\end{equation}

Then, the site-site correlation matrix $\bm{K}$ can be represented as an $N_\sigma \times N_\sigma$ symmetric matrix, which is $\bm{K} = C_0 \bm{A} \cdot \bm{A}^T$, and its trace is $\mathrm{Tr}(\bm{K}) = \sum_{i=1}^{N_\sigma} \bm{K}_{ii} = C_0$. So $C_0$ can be considered as a normalized factor. For such a symmetric matrix, $\bm{K}$ has $N_\sigma$ eigenvectors, $\bm{u}_J$, which are the eigen microstates of the system, where the $J = 1, 2, \cdots, N_\sigma$. It should be noetd that those eigen microstates can present the mesoscopic structure of the fluid system. These microstates can be represented as a unitary matrix, $\bm{U} = [\bm{u}_1, \bm{u}_2, \cdots, \bm{u}_{N_\sigma}]$.

By using the same procedure, an $M\times M$ matrix can be constructed to examine the correlation between the microstates, which is given by, $C=C_0 \bm{A}^T\cdot\bm{A}$, and similarly, the eigenvectors of $C$, $\bm{V}_I$, can be constructed as an $M \times M$ unitary matrix is $\bm{V} = [\bm{v}_1, \bm{v}_2, \cdots, \bm{v}_M]$.

Utilizing the singular value decomposition (SVD), the ensemble matrix $\bm{A}$ can be expressed as,

\begin{equation}
    \bm{A} = \bm{U} \cdot \mathrm{\sum} \cdot \bm{V}^T,
\end{equation}
where $\mathrm{\sum}$ is an $N_\sigma \times M$ diagonal matrix with elements.
\begin{equation}
    \mathrm{\sum}_{IJ} = 
    \begin{cases}
        \omega_I, & I=J \leq N_\sigma\\
        0, & \text{otherwise}
    \end{cases},
\end{equation}
where $\omega_I$ is the probability amplitudes of the microstate $I$.
Since the number of microstates exceeds the number of sites, only the first $N_\sigma$ vectors of $\bm{V}$ are required to reconstruct the ensemble matrix, 
\begin{equation}
    \bm{A} = \sum_{I=1}^{N_\sigma}\omega_I \bm{u}_I \otimes \bm{v}_I,
\end{equation}
 The microstate can be represented as,
\begin{equation}
    \delta \rho (j) = \sum_{I=1}^{N_\sigma}\omega_I\bm{u}_I\bm{v}_{jI}\sqrt{C_0}.
\end{equation}
Since $\mathrm{Tr}(\bm{K})=C_0$, we have $\sum_{I=1}^{N_\sigma} \omega^2_I=1$. Due to the translational and rotational symmetries of the system, the eigen microstate exhibits a distinct degeneracy structure.
For simplicity, we denote the merged probability amplitudes as $\lambda_I$ (ordered by magnitude), where each degenerate class corresponds to a single $\lambda_i$. Since the degenerate constitute a physically equivalent class, their spatial structure is collectively represented by $U_I$.

In the limit $L\to \infty$, $N\to \infty$, $M \to \infty$ at constant density $\rho$, all probability amplitudes drop to zero in the ensemble without localization of the microstate; conversely, if a probability amplitude converges to a finite value, it indicates that the corresponding eigen microstate undergoes condensation~\cite{Hu2019}.
Therefore, the probability amplitude, $\lambda_I$, functions as the order parameter for the phase transition. The corresponding eigen microstate, $U_I$, describes the mesoscopic structure of the emerged phase.

\subsection{\label{sec:2.2}Phase transition and finite-size scaling}
\begin{figure}[htp]
\center
\includegraphics[width=7cm]{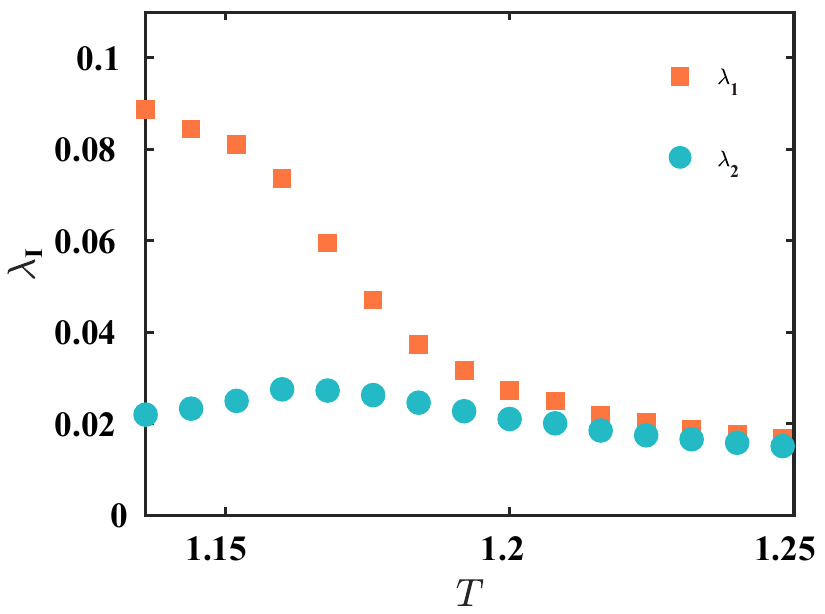}
\caption{The first two probability amplitudes, $\lambda_1$ and $\lambda_2$, near the critical region at the largest simulation scale $L_\sigma = 35$. The error bars are smaller than the symbol size.}
\label{fig:S1_S2L35}
\end{figure}
For the LJ fluid system, it contains both continuous and first-order phase transition. 
In this study, we focus on the continuous phase transition.
The continuous phase transition occurs at the critical temperature $T_c$ and critical density $\rho_c$. 
Notably, around the critical point, thermodynamic properties such as the correlation length experience a divergence, indicating strongly correlated fluctuations throughout the system. The variations in order parameter, correlation length, susceptibility, among others, with respect to $t = (T - T_c)/T_c$ are governed by a consistent set of critical exponents within the same universality class, i.e., three-dimensional Ising universality class. 

The finite-size effects can be pronounced for systems of limited size~\cite{privman1990finite}. For instance, the correlation length might be restricted to the size $L_\sigma$ near the critical point, leading to potential inaccuracies in determining the critical temperature based solely on a single system size. To counter this, a well-established finite-size scaling approach is employed, drawing inferences from multiple system sizes to discern and validate the critical temperature and its universality class.

Considering the EMT to continuous phase transition, it's ascertained that a specific finite-size scaling behavior in terms of system size $L_\sigma$.
In the vicinity of the critical point ($T_c$, $\rho_c$), the probability amplitude $\lambda_I$ of an eigen microstate follows the finite-size scaling form,
\begin{equation}
    \label{LTrho}
    \lambda_I(T,\rho,L_\sigma)=L_\sigma^{-\beta/\nu} F_I(tL_\sigma^{1/\nu}, \Delta \rho L_\sigma^{1/\nu'}),
\end{equation}
where $t = (T - T_c)/T_c$, $\Delta \rho = (\rho - \rho_c)/\rho_c$, $\nu$ and $\nu'$ are the critical exponent for diverging correlation length at $\Delta \rho = 0$ and $t=0$, respectively. $\beta$ is the order-parameter's critical exponent.
To determine the critical temperature and density, the preceding equation is reformulated as,
\begin{equation}
    \ln \lambda_I(T, \rho, L_\sigma) = -(\beta/\nu) \ln L_\sigma + \ln F_I(tL_\sigma^{1/\nu}, \Delta \rho L_\sigma^{1/\nu'}).
    \label{LTrholog}
\end{equation}
A linear relationship between $\ln \lambda_I$ and $\ln L_\sigma$ arises only at the critical point ($\rho = \rho_c$ and $T=T_c$). When the system is away from the critical potint, the relationship deviates from linearity. This scaling behavior enables the  determination of the critical temperature and density.
Eq.~(\ref{LTrho}) leads to the definition of a useful dimensional ratio,
\begin{equation}
    R(T, \rho, L_\sigma) = \frac{\lambda_2 (T, \rho, L_\sigma)}{\lambda_1 (T, \rho, L_\sigma)} = \frac{F_2(tL_\sigma^{1/\nu}, \Delta \rho L_\sigma^{1/\nu'})}{F_1(tL_\sigma^{1/\nu}, \Delta \rho L_\sigma^{1/\nu'})}.
\end{equation}
The ratio $R(T, \rho, L_\sigma)$ depends on the reduced temperature $t$, the reduced density $\Delta \rho$, and the system size $L_\sigma$.
It indicates that the ratio of two probability amplitudes is independent of the system size, $L_\sigma$, at the critical point, which is a robust metric to pinpoint the critical temperature, $T_c$, and critical density, $\rho_c$.

In this work, the first eigen microstate, $\lambda_1$, corresponds to a six-fold degenerate state, while the second eigen microstate, $\lambda_2$, is 12-fold degenerate.

After obtain the crtical point, we try to calculate the critical exponents.
When $\rho = \rho_c$ and $T=T_c$, the scaling form Eq.~(\ref{LTrho}) reduces as,
\begin{equation}
    \ln \lambda_I(T_c, \rho_c, L_\sigma) = -(\beta/\nu) \ln L_\sigma + c_1,
\label{loglog}
\end{equation}

where $c_1$ is a constant. The above equation shows that at the critical point, there is a linear relation between $\ln L_\sigma$ and $ \ln \lambda_I $, and the slope of the line is the exponent ratio $ \beta/\nu $.
This relation can be used to determine the critical exponent ratio $ \beta/\nu $.

\begin{figure*}[htp]
    \center
    \includegraphics[width=16cm]
    {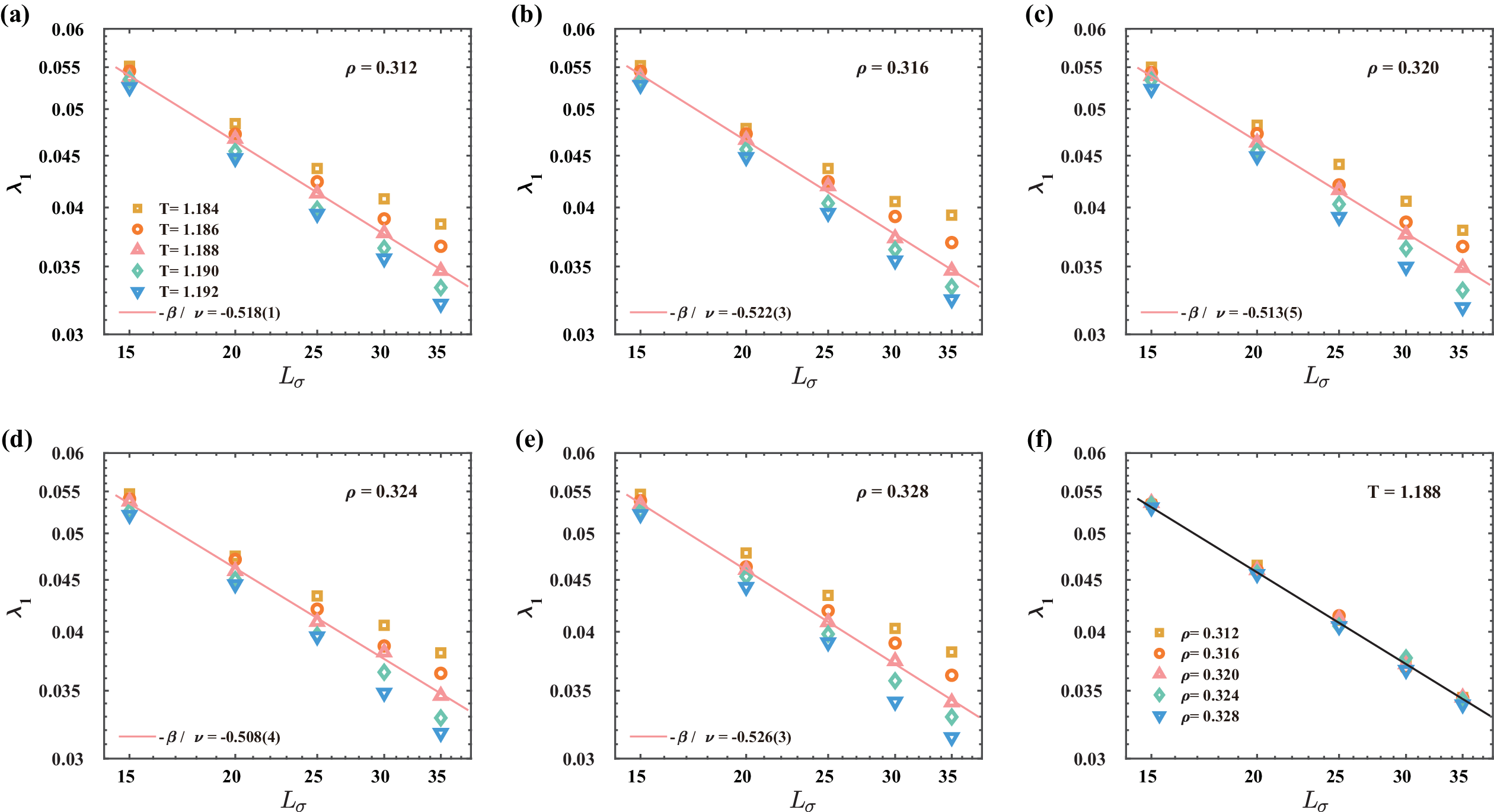}
    \caption{(a-e) Log-log plot of $\lambda_1$ against system size $L_\sigma$ at different target densities: (a) $\rho=0.312$, (b) $\rho=0.316$, (c) $\rho=0.320$, (d) $\rho=0.324$, (e) $\rho=0.328$. For each density, data are shown for a series of temperatures near the critical region. Across all densities, the double-logarithmic plots of $\lambda_1$ versus system size $L_\sigma$ exhibit the most robust linear behavior at $T=1.188$ (solid lines), compared to the clear deviations observed at other temperatures.
    (f) fixes the temperature at $T=1.188$ and compares different densities. At this stage, it is not possible to effectively distinguish the critical density.}
    \label{fig:get_tc}
\end{figure*}
\begin{figure*}[htp]
    \center
    \includegraphics[width=16cm]
    {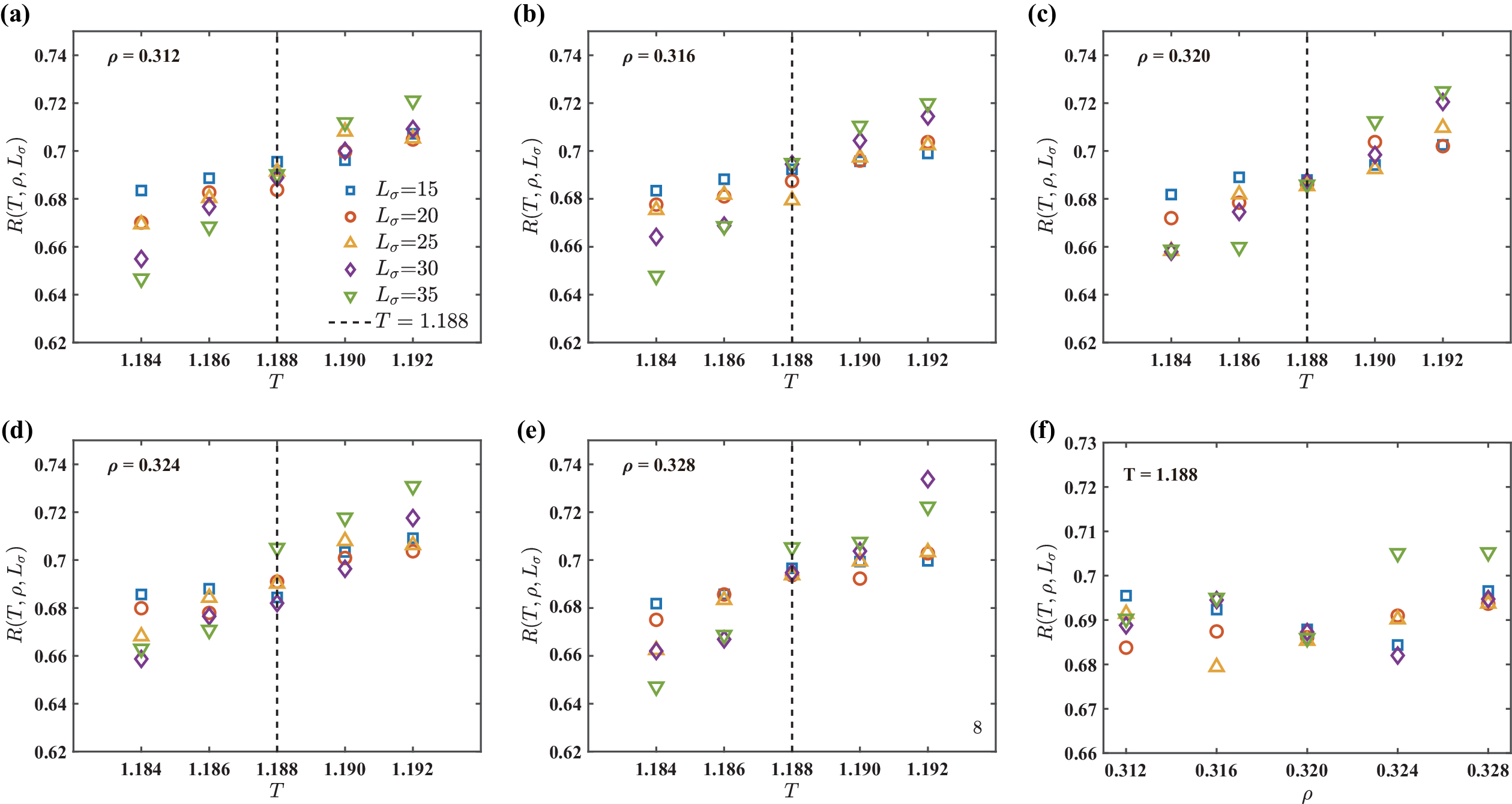}
    \caption{Determination of the critical point through finite-size scaling of probability amplitude ratios. (a-e) The ratio of the first two probability amplitudes $R(T, \rho, L_\sigma)= \lambda_2/\lambda_1$, as a function of temperature for different system sizes $L_\sigma$ at densities: (a) $\rho=0.312$, (b) $\rho=0.316$, (c) $\rho=0.320$, (d) $\rho=0.324$, (e) $\rho=0.328$. The curves for different $L_\sigma$ intersect at a common point only when both the density and temperature reach their critical values. (f) The ratio $R(T, \rho, L_\sigma)$ as a function at the fixed temperature $T=1.188$. The curves for different $L_\sigma$ intersect precisely at $\rho=0.320$, confirming the critical density $\rho_c=0.320(4)$. These consistent crossing behaviors collectively determine the critical point at $T_c=1.188(2), \rho_c=0.320(4)$.}
    \label{fig:get_rhoc}
\end{figure*}
Since when $\rho = \rho_c$, the ratio $R(T, \rho_c, L_\sigma)$ is can be reduced as,
\begin{equation}
    R(T,\rho_c, L_\sigma) = \frac{\lambda_2}{\lambda_1} = \frac{F_2(tL_\sigma^{1/\nu})}{F_1(tL_\sigma^{1/\nu})}.
    \label{eq:ratio}
\end{equation}
To determine the critical exponent $\nu$, the following equation is proposed,
\begin{equation}
    \ln \left.\frac{\partial R(T,\rho_c,L_\sigma)}{\partial t} \right|_{t\to 0} = (1/\nu) \cdot \ln L_\sigma + c_2.
    \label{eq:nu} 
\end{equation}
where $c_2$ is a constant. Then $\nu$ can be determined by fitting the slope of the above equation. The critical exponent $\nu'$ can similarly be obtained by fitting the slope of the following equation,
\begin{equation}
    \ln \left.\frac{\partial R(T_c,\rho,L_\sigma)}{\partial \Delta \rho} \right|_{\Delta \rho\to 0} = (1/\nu') \cdot \ln L_\sigma + c_3,
    \label{eq:nu'}
\end{equation}
where $c_3$ is a constant. This analysis is not pursued in the present work but will be addressed in the future studies.

\section{\label{sec:3}Results and discussion}
The canonical ensemble Monte Carlo simulation is used to generate the microstates of the LJ system.
The simulation is performed in a three-dimensional cubic box of size $L$.
We apply periodic boundary conditions in all directions. 
Recent advances in collective update techniques have greatly improved the efficiency of Monte Carlo simulations of fluids, especially through the clustering algorithm for hard spheres and its extension to LJ systems~\cite{Buhot1998,Liu2004}. 
Given the substantial particle count in this study, an efficient simulation method, rejection-free geometric cluster algorithm~\cite{Liu2004}, is adopted to generate the microstates of the LJ system.


As noted in previous studies~\cite{stephan2019thermophysical, smit1992phase}, the reported critical temperature and critical density of the LJ fluid exhibit considerable scatter. 
The critical density is known to depend on both the truncation radius, $R_c$, of the LJ potential and the treatment of its tail (e.g., shifted or unshifted)~\cite{stephan2018vapor}.
In this work, we adopt the same truncated LJ potential model as that used by N. B. Wilding, with a cutoff radius of $R_c = 2.5\sigma$ and an unshifted potential~\cite{wilding1995critical}.

Simulations were carried out in the canonical ensemble for system sizes $L_\sigma=15, 20, 25, 30, 35$. A high-resolution sweep of the $\rho-T$ plane was performed, with the particle number fixed at $N = \rho L^3$ for each density. For each state point, the first 80,000 Monte Carlo steps were discarded to ensure equilibration. Following equilibration, a statistical ensemble of $M = 10^5$ microstates was sampled at intervals of 100 Monte Carlo steps. These collected configurations were then encoded into the normalized ensemble matrix $\bm{A}$, from which the correlation matrix, probability amplitudes, and eigen microstates were derived.

\begin{table}[htbp]
  \caption{Critical properties of the LJ fluid: A comparison of different methods.(Note: MC = Monte Carlo simulation, MD = Molecular Dynamics simulation)}
  \label{tab:crit_props}
  \begin{ruledtabular}
  \begin{tabular}{llccccl}
    $R_c$ & Method & $\rho_c$ & $T_c$ & & Source \\
    \hline
    $2.5$ & MC($\mu$VT) & 0.3197(4) & 1.1876(3) & & Wilding~\cite{wilding1995critical} \\
    $2.5$ & MD(NVT) & 0.319 & 1.186 & & Trokhymchuk and Alejandre~\cite{trokhymchuk_1999} \\
    $2.5$ & MC(NVT) & 0.320(4) & 1.188(2) & & this work \\
  \end{tabular}
  \end{ruledtabular}
\end{table}

\begin{figure}[htp]
    \center
    \includegraphics[width=6cm]
    {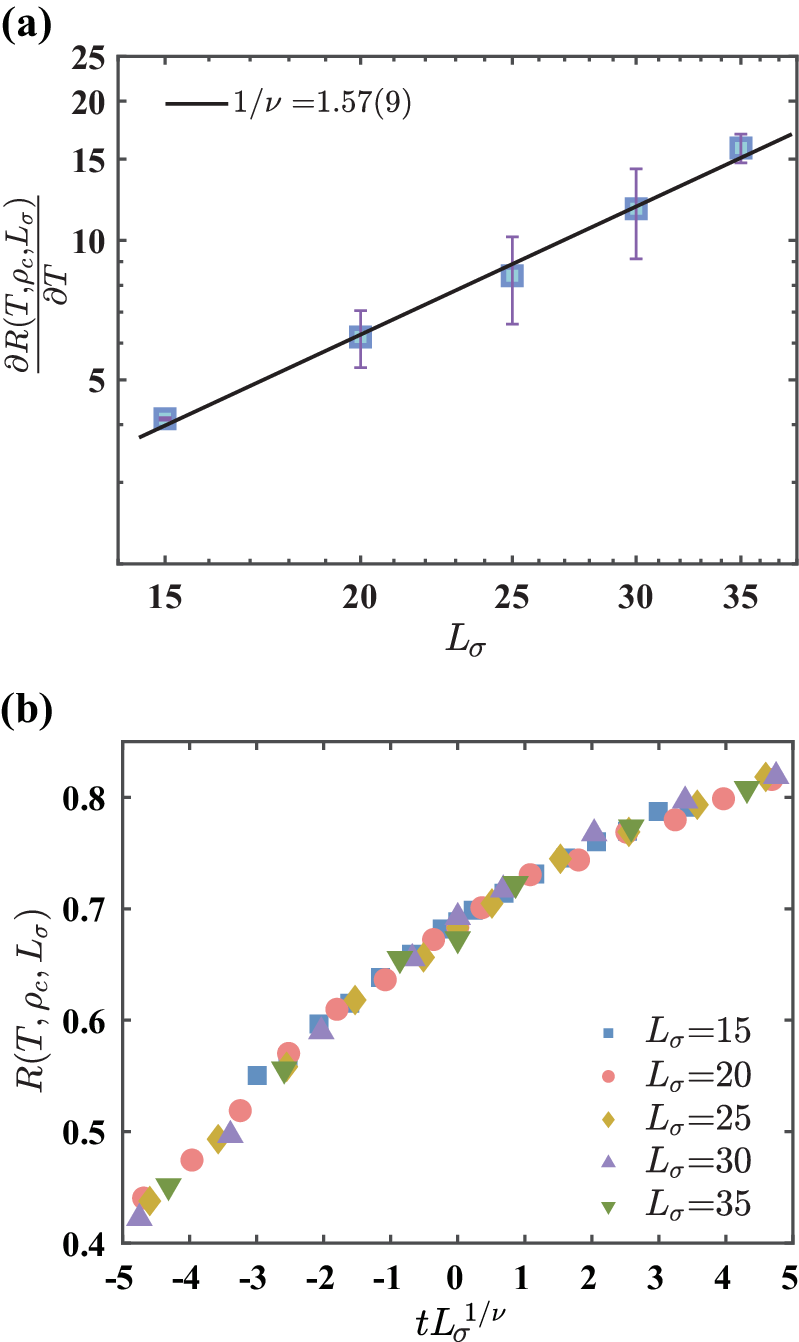}
    \caption{(a) According to Eq.~(\ref{eq:nu}), the slope of solid black line corresponds to $1/\nu$, where $\nu$ is the critical exponent associated with the correlation length. The fitted slope gives $\nu = 0.64(3)$. (b) Finite-size scaling form of $R(T, \rho_c, L_\sigma)$ when $\nu=0.64$.}
    \label{fig:nu_2v1_S}
\end{figure}

\begin{figure*}[htb]
    \center
    \includegraphics[width=11cm]
    {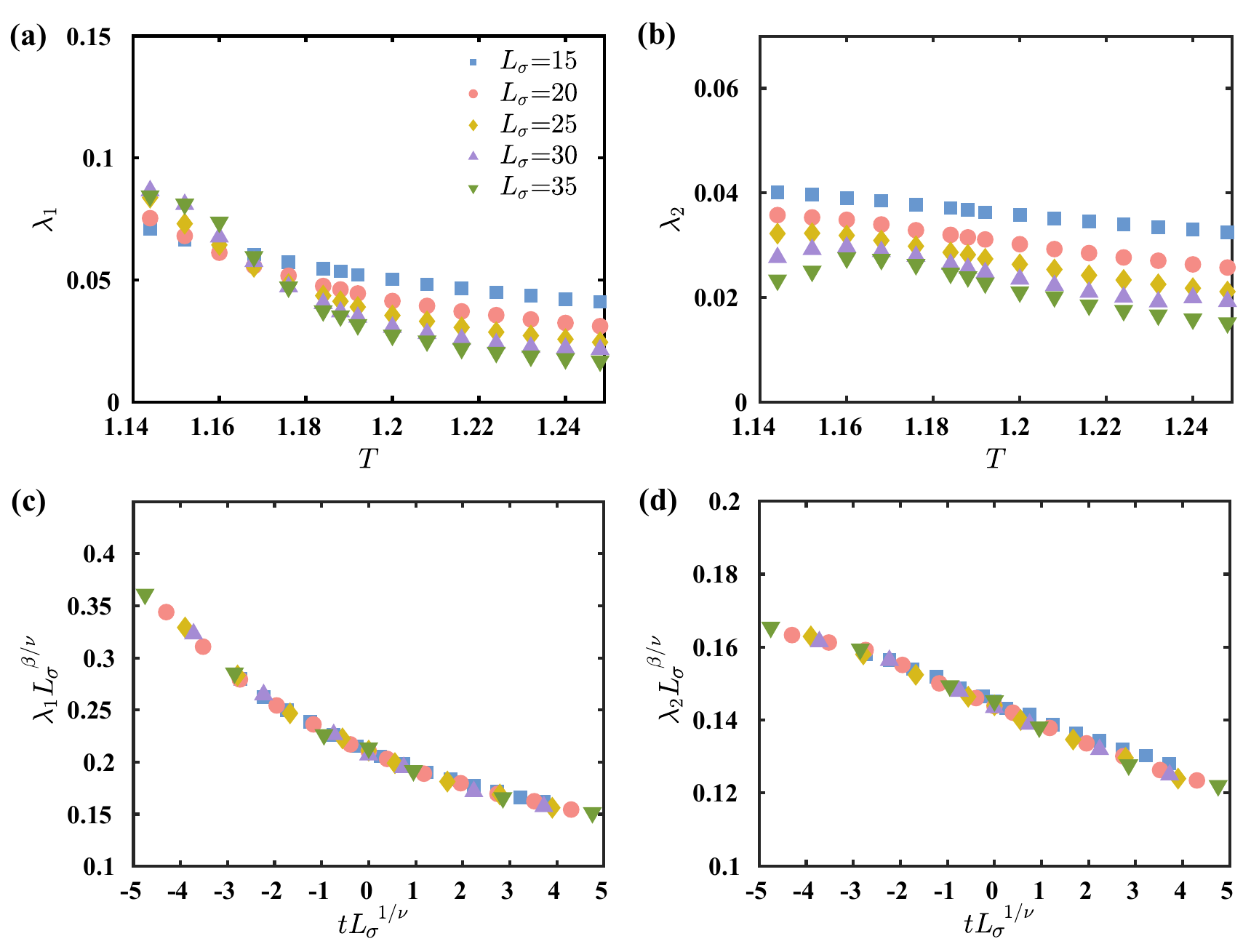}
    \caption{Scaling behavior of the first two probability amplitude near the critical point across different system sizes. (a) The first probability amplitude $\lambda_1$ and (b) its finite-size scaling form by $\nu=0.64$ and $\beta / \nu = 0.513$. (c) Shows the second probability amplitude $\lambda_2$ and (d) its finite-size scaling form by the same critical exponents.}
    \label{fig:S1and7_S}
\end{figure*}

The behavior of the derived probability amplitudes near criticality is displayed in Fig.~\ref{fig:S1_S2L35}. Data for the largest system size ($L_\sigma=35$) at $\rho=0.32$ reveal the sensitivity of the two leading amplitudes to the temperature changes.

At high temperatures, all probability amplitudes have comparable magnitudes, consistent with a disordered gas-like phase. As the temperature decreases, the first amplitude $\lambda_1$ grows rapidly and becomes dominant, signaling the emergence of an ordered liquid-like structure.
The second amplitude $\lambda_2$ exhibits a slight initial rise followed by a decrease as $\lambda_1$ becomes dominant. This behavior indicates the suppression of subleading modes as the system approaches the critical region. The degeneracies of these modes -- six fold for $\lambda_1$ and twelve-fold for $\lambda_2$ -- are consistent with the cubic symmetry of the simulation model.

The evolution of probability amplitudes provides initial evidence of the phase transition. To precisely locate the critical point and extract quantitative critical exponents, we perform a finite-size scaling analysis based on Eq.~(\ref{LTrholog}). 
This analysis employs simulations conducted at multiple densities ($\rho = 0.312$ to $0.328$, in steps of $0.004$). 
For each density, probability amplitudes were computed for different system sizes at five temperatures in the critical region ($T = 1.184$ to $1.192$ in increments of $0.002$).
As shown in FIG.~\ref{fig:get_tc}(a)-(e), the double-logarithmic plots of $\lambda_1$ versus $L$ exhibit a well-defined linear relationship at $T = 1.188$ across all densities, consistent with the power-law scaling expected at criticality. Deviations from this linearity are observed at other temperatures, $\lambda_1$ curves upward with increasing $L$ for $T < 1.188$ and downward for $T > 1.188$.

The scaling behavior at $T = 1.188$ is further examined in Fig.~\ref{fig:get_tc}(f), which consolidates data across all studied densities. While a robust power-law dependence is evident, the data for different densities exhibit similar linear trends, making it impossible to effectively distinguish the critical density solely from $\lambda_1$. This limitation necessitates a more stringent criterion, according to Eq.~(\ref{eq:ratio}), a fixed-point analysis of the ratio $R = \lambda_2/\lambda_1$, whose size-independent crossing behavior is expected to uniquely determine the critical parameters.

Figure~\ref{fig:get_rhoc}(a)-(e) present the evolution of $R$ across different system sizes, revealing a clear and unique crossing behavior only when the system simultaneously approaches the critical density and critical temperature. 
Specifically, based on the precise intersection of curves at the fixed temperature in FIG.~\ref{fig:get_rhoc}(f), we pinpoint the critical density as $\rho_c = 0.320(4)$, and the critical temperature as $T_c = 1.188(2)$. 
Concurrently, the robust power-law scaling shown in FIG.~\ref{fig:get_tc}(c) confirms the critical temperature as $T_c = 1.188(2)$. 
The critical parameters obtained in this work are consistent within uncertainty with previously reported values, as summarized in Table~\ref{tab:crit_props}.

\begin{figure*}[hbtp]
    \center
    \includegraphics[width=0.9\textwidth]{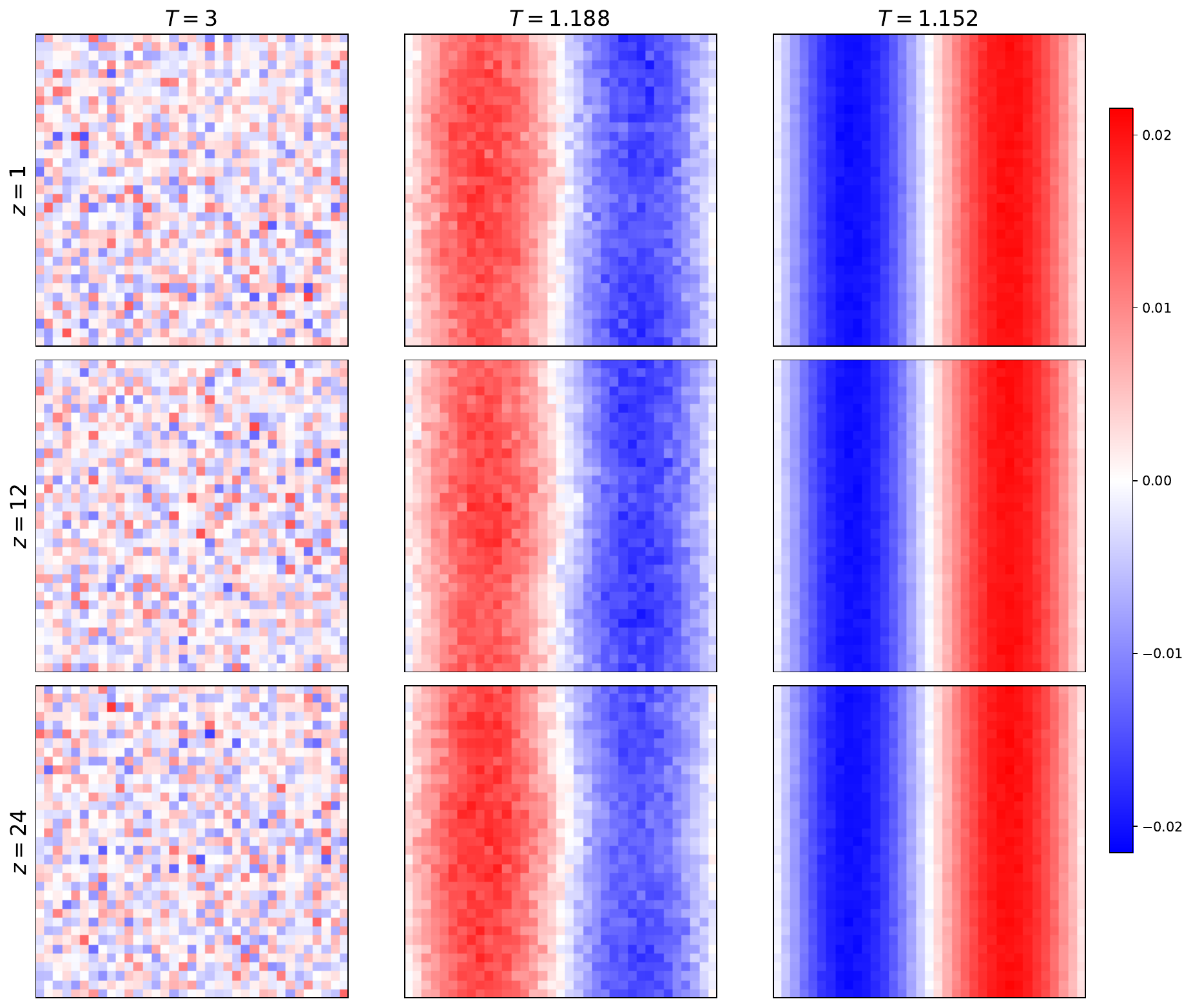}
    \caption{Structure of the first eigen microstate at critical density for different temperatures ($T=1.152, 1.188$, and $3.0$). The system size is fixed at $L_\sigma = 35$.}
    \label{fig:U1}
\end{figure*}

\begin{figure*}
    \center
    \includegraphics[width=0.9\textwidth]{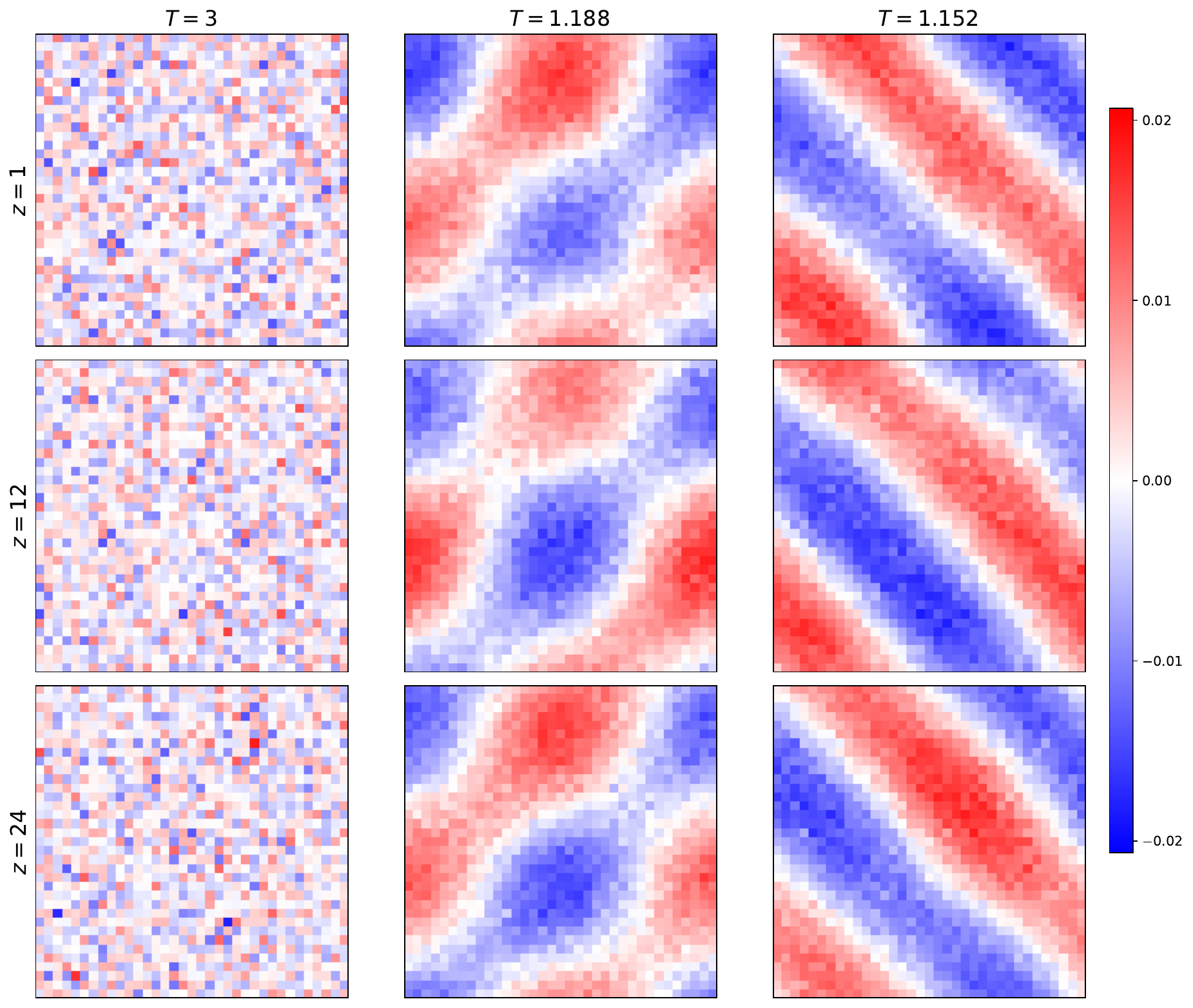}
    \caption{Structure of the second eigen microstate at critical density for different temperatures ($T=1.152, 1.188$, and $3.0$). The system size is fixed at $L_\sigma = 35$.}
    \label{fig:U1_2}
\end{figure*}

The established critical point enables reliable extraction of critical exponents through finite-size scaling analysis. The validity of this scaling analysis is contingent upon the system being at the critical point. Therefore, the slope of the fitted line in FIG.~\ref{fig:get_tc}(c) directly yields the critical exponent ratio. By calculating this slope according to Eq.~(\ref{loglog}), we obtain $\beta/\nu=0.513(5)$. This result is in agreement with Wilding~\cite{wilding1995critical} within the error margin.

To determine the individual values of the critical exponents beyond their ratio, we analyze the scaling behavior using Eq.~(\ref{eq:nu}). The slope in FIG.~\ref{fig:nu_2v1_S}(a) corresponds to $1/\nu$, where $\nu$ is the correlation length exponent. From this relation, we obtain $\nu = 0.64(3)$. This value agrees within uncertainty with $\nu = 0.63(4)$ reported in Ref.~\onlinecite{watanabe2012a} and is consistent with the critical exponent of the three-dimensional Ising universality class~\cite{hasenbusch2010}.Having determined the exponent $\nu$, we now examine the finite-size scaling of $R(T, \rho, L_\sigma) = \lambda_2/\lambda_1$ using this result. As shown in FIG.~\ref{fig:nu_2v1_S}(b), the data for different system sizes collapse onto a universal curve under this scaling analysis.

Furthermore, using the ratio $\beta/\nu = 0.513(5)$ and the determined value of $\nu$, we calculate the order-parameter exponent $\beta = 0.32(2)$. This result is consistent with the value $\beta = 0.3269(6)$ reported for the three-dimensional Ising model on a simple cubic lattice~\cite{talapov_magnetization_1996}.Based on the critical exponents $\nu$ and $\beta$ determined above, we further examine the scaling behavior of the probability amplitudes near the critical point within a more stringent scaling framework. Compared to the previous analysis that relied solely on $\nu$, the simultaneous incorporation of both $\nu$ and $\beta$ imposes stronger constraints on the scaling transformation. This approach provides a more rigorous test for validating the critical exponents and confirming the universality class of the phase transition.

Figure~\ref{fig:S1and7_S}(a) and (b) show the first two probability amplitudes, $\lambda_1$ and $\lambda_2$, for all system sizes. Their scaled forms, $\lambda_I L_\sigma^{\beta/\nu}$, are plotted as functions of $tL_\sigma^{1/\nu}$ in FIG.~\ref{fig:S1and7_S}(c) and (d), where $\beta/\nu = 0.513$ and $\nu = 0.64$ as derived previously. The clear data collapse across all system sizes under this more stringent scaling framework confirms the robustness of the finite-size scaling behavior near criticality. This result provides further validation of the accuracy of the critical exponents and reinforces the consistency of the phase transition with the three-dimensional Ising universality class.

An investigation of the spatial structure of the first eigen microstates at critical density across different temperatures complements the analysis of critical behavior. 
FIG.~\ref{fig:U1} displays the cross-sectional slices of the first eigen microstate, $U_1$. Slices are taken at $z = 1, 12, 24$ for three characteristic temperatures, $T=1.152$ (below $T_c$), $T=1.188$ (at $T_c$), and $T=3.0$ (above $T_c$), with system size $L_\sigma=35$.
At high temperatures, $U_1$ exhibits no discernible structure, consistent with a disordered gas phase. 
As the temperature decreases, layer-like phase separation emerges, characterized by regions of higher density (red, representing the liquid phase) and lower density (blue, representing the gas phase) compared to the average density.
Pronounced density fluctuations appear at the critical temperature.
With further cooling, these fluctuations subside, allowing for a clear, sharp-boundary phase separation between the two clusters.
This progression visually illustrates the gradual formation of the liquid phase during the phase transition.
It should be noted that these mesoscopic spatial structures can not be captured from the raw microstates, illustrating the advantage of EMT.

We further examine the spatial structures of the second eigen microstates, $U_2$, across different temperatures at the critical density, as shown in FIG.~\ref{fig:U1_2}.
In high temperature, $U_2$, also displays no discernible structure, consistent with a disordered gas-like phase.
As the temperature decreases, the second eigen microstate exhibits a four-cluster structure within the $xy$-plane, a pattern that remains consistent across different $z$-coordinates.
As the temperature decreases further, clusters of the same phase tend to merge, forming larger domains.

Collectively, these findings confirm that the continuous phase transition in the LJ system falls within the Ising universality class, and demonstrate the effectiveness of EMT in characterizing liquid-gas phase transitions.

\section{Conclusion}
\label{sec:conclusion}
In this work, we have applied the EMT to characterize the liquid-gas critical behavior of the LJ fluid in the canonical ensemble. By combining EMT with finite-size scaling, we have simultaneously determined the critical temperature and density,$T_c = 1.188(2)$ and $\rho_c = 0.320(4)$. And extracted the associated critical exponents, $\beta = 0.32(2)$ and $\nu = 0.64(3)$. These values confirm that the transition belongs to the three-dimensional Ising universality class.

A key advantage of employing EMT to study the continuous phase transition of the LJ system is its ability to  determine the critical temperature and density without requiring a priori knowledge, such as a predefined order parameter or the law of rectilinear~\cite{watanabe2012a}.
In contrast, EMT constructs the ensemble matrix directly from particle-based density fluctuations. 
This approach offers a more straightforward and physically transparent route to locating the critical region while maintaining high accuracy. 
As shown in FIG.~\ref{fig:get_rhoc}, both the critical density and temperature are obtained simultaneously by analyzing the behavior of $R(T, \rho, L_\sigma)$ at various temperatures and densities. 
Importantly, eigen microstates disclose mesoscopic spatial structures near the critical region, offering insight that raw microstates do not provide.
Furthermore, the use of the NVT ensemble ensures precise control over both temperature and particles number throughout the simulation.
In this work, the site size is set equal to $\sigma$, which is the characteristic length parameter of the LJ fluid. While local fluctuations at this scale inlude discreteness effects, EMT isolates the spatially correlated, critical fluctuations into the leading eigen microstates, therby separating the macroscopic physical signal from the uncorrelated local noise.

This flexibility opens avenues for investigating complex molecular systems, particularly those involving interplay between critical and phase-separation transitions where order parameters are not well defined. Realizing this potential, however, relies on the judicious choice of microstates that capture the essential order-disorder characteristics of the transition.

Despite the critical importance of this choice, the EMT has demonstrated broad applicability from the Ising model to Earth sciences. This track record supports its potential extension to more complex systems, such as confined fluids and molecular mixtures.

\begin{acknowledgments}
This work is supported by the National Natural Science Foundation of China (Grant No. 12135003, 12405032) and the Fundamental Research Funds for the Central Universities.
\end{acknowledgments}

\nocite{*}
\bibliography{reference}

\end{document}